\documentclass[aps,prc,preprint,showpacs,superscriptaddress]{revtex4}  
\usepackage{graphicx}
\usepackage{amsmath}
\newcommand {\nc} {\newcommand}
\nc {\beq} {\begin{eqnarray}} \nc {\eol} {\nonumber \\} \nc {\eeq}
{\end{eqnarray}} \nc {\eeqn} [1] {\label{#1} \end{eqnarray}} \nc
{\eoln} [1] {\label{#1} \\} \nc {\ve} [1] {\mbox{\boldmath $#1$}}
\nc {\rref} [1] {(\ref{#1})} \nc {\Eq} [1] {Eq.~(\ref{#1})} \nc
{\re} [1] {Ref.~\cite{#1}} \nc {\dem} {\mbox{$\frac{1}{2}$}} \nc
{\arrow} [2] {\mbox{$\mathop{\rightarrow}\limits_{#1 \rightarrow
#2}$}}
\begin{document}
\title{Theoretical study of the $\alpha+d$ $\rightarrow$ $^6$Li + $\gamma $ astrophysical capture process in a three-body model}

\author {E. M. Tursunov}
\email{tursune@inp.uz} \affiliation {Institute of Nuclear Physics,
Academy of Sciences, 100214, Ulugbek, Tashkent, Uzbekistan}
\author {A. S. Kadyrov}
\email{a.kadyrov@curtin.edu.au} \affiliation{Department of Physics
and Astronomy, Curtin University, GPO Box U1987, Perth, WA 6845,
Australia}
\author {S. A. Turakulov}
\email{turakulov@inp.uz} \affiliation {Institute of Nuclear Physics,
Academy of Sciences, 100214, Ulugbek, Tashkent, Uzbekistan}
\author {I. Bray}
\email{i.bray@curtin.edu.au} \affiliation{Department of Physics and
Astronomy, Curtin University, GPO Box U1987, Perth, WA 6845,
Australia}

\begin{abstract}
The astrophysical capture process  $\alpha+d$ $\rightarrow$ $^6$Li +
$\gamma$ is studied in a three-body model.
 The initial state is factorized into the deuteron bound state and the $\alpha+d$ scattering state. The final nucleus
 $^6$Li(1+)  is described as a three-body bound state $\alpha+n+p$ in the hyperspherical Lagrange-mesh method.
 The contribution of the E1 transition operator from the initial isosinglet states to the isotriplet components
 of the final state is estimated to be negligible.  An estimation of the forbidden E1 transition to the isosinglet
 components of the final state is comparable with the corresponding results of the two-body model.
 However, the contribution of the E2 transition operator is found to be much smaller than the corresponding estimations of the two-body model.
 The three-body model perfectly matches the new experimental data of the LUNA
 collaboration with the spectroscopic factor 2.586 estimated from the
 bound-state wave functions of $^6$Li and deuteron.
 \end{abstract}


 \maketitle

\section{Introduction}
\par In the Big Bang nucleosynthesis (BBN) model of the Universe estimations of
the primordial abundance of the light  $^2$H, $^{3}$He and  $^{4}$He
nuclei are in very good agreement with astrophysical observations
\cite{fields11}. However, the situation is very different for
 the primordial abundance of the $^{6}$Li and $^{7}$Li nuclei \cite{spite82,asp06,asp08,lind13,stef12}.
 Recent observations of $^6$Li in metal-poor stars \cite{asp06} suggest a large production of this isotope.
The data for the $^6$Li/$^7$Li ratio of  about 0.05 is almost three
orders of magnitude larger than estimations
 from the BBN model \cite{serp04}.  Understanding of this phenomenon  is one of the open problems
 in nuclear astrophysics.

In BBN the light $^6$Li nucleus is produced   mainly through the
radiative capture process
\begin{eqnarray} \label{1}
\alpha+d\rightarrow^6\mathrm{Li}+\gamma
\end{eqnarray}
at low energies within the range $50 \le E_{\rm cm} \le 400$ keV
\cite{serp04}. This process was experimentally studied in detail  at
energies around the $3^+$  resonance of $E_{\rm cm}=$0.711 MeV and
above \cite{mohr94,robe81}. Until recently the direct measurement of
the cross section of the process at low energies was not possible
due to serious experimental difficulties \cite{kien91,hamm10}. In
Ref. \cite{hamm10} breakup  of the $^6$Li nucleus in the field of
heavy ion $^{208}$Pb was studied with the aim to extract data on the
cross section of the inverse process at astrophysical energies in
laboratory conditions. However, dominance of the nuclear breakup
over the Coulomb induced process did not allow to implement this
idea. The LUNA collaboration has recently reported  new data at two
astrophysical energies E=94 keV and E=134 keV  \cite{luna14}.  The
results turn out to be much lower than the old data from Ref.
\cite{kien91}. Recently in Ref.\cite{MSB16} a way to improve the
accuracy of the direct experiment has been proposed based on the
photon angular distribution calculated in the potential model. The
results provide the best kinematic conditions for the measurement of
the $^2$H($\alpha,\gamma$)$^6$Li reaction.

 From the theoretical side, different two-body and three-body potential models
\cite{lang86,dub951,dub952,type97,desc98,mukh11,sade15,tur15} and
{\em ab initio} approaches \cite{noll01} have been developed. These
studies have demonstrated that the main contribution to the process
at energies around and beyond the $3^+$ resonance comes from the E2
transition. However, at low astrophysical energies the situation is
different. Here the dominant contribution comes from the
E1-transition operator. The most realistic two-body model of
Ref.\cite{mukh11} is based on the well-known asymptotic form of the
two-body $\alpha+d$  bound-state wave function at low energies and a
complicated potential derived from the original Woods-Saxon
potential via the integro-differential transformation at higher
energies. Recently these results have been reproduced with a much
simpler $\alpha-d$ potential of the Gaussian form describing both
bound state (ANC, binding energy) and scattering state (phase shifts
in the $S$, $P$, $D$-waves) properties \cite{tur15} of the
$\alpha+d$ system.

On the other hand, in the two-body models the E1 transition is
forbidden by the isospin-selection rule, since both initial and
final states are isospin singlet.
  To overcome this problem, an appropriate correction to the
E1-transition operator was introduced to take into account the
difference between the mass of the alpha-particle and the twice the
deuteron mass. Without this correction the E1 transition does not
contribute to the S-factor of the process. However, this drawback
has been common for all the models developed so far.

There is another possible development for the estimation of the E1-
and E2- transition matrix elements for the $^4$He$(d,\gamma)^6$Li
capture process. In realistic three-body models the E1 transition is
allowed from the initial $T_i=0$  states  to the $T_f=1$ components
of the final $^6$Li$(1^+)$  bound state of the $\alpha+n+p$ system.
Indeed, the ground state of the $^6$Li nucleus contains a small
isospin-triplet component.  The norm square of this component of the
three-body wave function in hyperspherical coordinates
\cite{desc03,tur06} is about 1.13 $\times 10^{-5}$. However, it
still can make some additional contribution to the process.

The aim of present study is to estimate the E1- and E2-transition
contribution to the S-factor of the afore-mentioned process in a
three-body model.  The initial three-body
 wave function is factorized into the deuteron bound-state and the $\alpha+d$ scattering wave functions. The
final $^6$Li(1+) state is described as a $\alpha+p+n$ three-body
bound system. The hyperspherical wave function on the Lagrange mesh
basis available for the $^6$Li(1+) bound-state \cite{desc03,tur06}
will be used.

In section 2 we describe the model, in section 3 we discuss obtained
numerical results and finally, in the last section we make
conclusions.

\section{Theoretical model}
\subsection{Cross sections of the radiation capture process}

The cross sections of the radiative capture process reads
\begin{align}
\sigma_{E}(\lambda)=& \sum_{J_i T_i \pi_i}\sum_{J_f T_f
\pi_f}\sum_{\Omega \lambda}\frac{(2J_f+1)} {\left [I_1
\right]\left[I_2\right]} \frac{32 \pi^2 (\lambda+1)}{\hbar \lambda
\left( \left[ \lambda \right]!! \right)^2} k_{\gamma}^{2 \lambda+1}
C^2_S \nonumber \\ &\times \sum_{l_\omega I_\omega}
 \frac{1}{ k_\omega^2 v_\omega}\mid
 \langle \Psi^{J_f T_f \pi_f}
\|M_\lambda^\Omega\| \Psi_{l_\omega I_\omega}^{J_i T_i \pi_i}
\rangle \mid^2,
\end{align}
where $\Omega=$E  or M (electric or magnetic transition), $\omega$
denotes the entrance channel, $k_{\omega}$, $v_\omega$,  $I_\omega$
are the wave number, velocity of the $\alpha-d$ relative motion and
the spin of the entrance channel, respectively, $J_f$, $T_f$,
$\pi_f$ ($J_i$, $T_i$, $\pi_i$) are the spin, isospin and parity of
the final (initial) state, $I_1$, $I_2$ are channel spins,
$k_{\gamma}=E_\gamma / \hbar c$ is the wave number of the photon
corresponding to the energy $E_\gamma=E_{\rm th}+E$ with the
threshold energy $E_{\rm th}=1.474$ MeV. The wave functions
$\Psi_{l_\omega I_\omega}^{J_i T_i \pi_i} $ and $\Psi^{J_f T_f
\pi_f} $ present  the initial and final states, respectively. They
are given in a common form for the both two-body and three-body
models. The reduced matrix elements are evaluated between the
initial and final states. The constant $C^2_S$ is the spectroscopic
factor \cite{Angulo}. We also use short-hand notations $[I]=2I+1$
and $[\lambda]!!=(2\lambda+1)!!$.

The electric-transition operator in the Jacobi coordinates can be
written as \cite{desc03}
\begin{align}
M_{\lambda \mu}^{E}(\vec{x},\vec{y}) =&
e\left[ \hat{Z}_{12} \left( \frac{-A_3}{A}
\right)^{\lambda}+\hat{Z}_3 \left(\frac{A_{12}}{A} \right)^{\lambda}
\right]M_{\lambda \mu}^{E}(\vec{y})
\nonumber \\
&+e\left[\hat{Z}_1 \left( \frac{-A_2}{A_{12}}
\right)^{\lambda}+\hat{Z}_2 \left(\frac{A_1}{A_{12}}
\right)^{\lambda} \right]M_{\lambda \mu}^{E}(\vec{x})+
\nonumber\\
&+e\sum_{k>0}^{\lambda - 1}\alpha_{\lambda
k}\left(\frac{-A_3}{A}\right)^k \left[\hat{Z}_1 \left(
\frac{-A_2}{A_{12}} \right)^{\lambda-k} \right.
\nonumber \\
&+\hat{Z}_2 \left. \left(\frac{A_1}{A_{12}} \right)^{\lambda-k}
\right]
\left\{ M_{k}^{E}(\vec{y})\otimes M_{\lambda-k}^{E}(\vec{x})
\right\} _{\lambda \mu},
 \label{eq11}
\end{align}
with
\begin{eqnarray}
M_{\lambda \mu}^{E}(\vec{x})=\left( \frac{x}{\sqrt{\mu_{12}}}
\right) ^\lambda Y_{\lambda \mu} (\hat{x}) \equiv r^{\lambda}
Y_{\lambda \mu} (\hat{r}),
\end{eqnarray}
\begin{eqnarray}
M_{\lambda \mu}^{E}(\vec{y}) = \left( \frac{y}{\sqrt{\mu_{12}}}
\right) ^\lambda Y_{\lambda \mu} (\hat{y})\equiv R^{\lambda}
Y_{\lambda \mu} (\hat{R}),
\end{eqnarray}
and \beq \alpha_{\lambda k}=\left(\frac{4\pi [\lambda]!}{[k]!
[\lambda-k]!}\right)^{1/2},
 \eeq
where  \quad $\frac{1}{\mu_{12}}=\frac{1}{A_1}+\frac{1}{A_2}$ \quad
and  \quad $\frac{1}{\mu_{(12)3}}=\frac{1}{A_{12}}+\frac{1}{A_3}$
\quad are the reduced masses. The Jacobi coordinates $\ve{x}$
(between the proton and neutron), $\ve{y}$ (between the $p+n$ and
the $\alpha$-particle) and relative $\ve{r}$, $\ve{R}$ coordinates
are related as
 \beq \ve{x} = \sqrt{\mu_{12}}
\ve{r},
\quad \ve{y} = \sqrt{\mu_{(12)3}} \ve{R}. \eeq

\subsection{Wave functions}

In the present three-body model the initial state is factorized as
\begin{align}
\Psi_{i}^{J'M',T'0}(\vec{x},\vec{y})=&\frac{u_{l'}^{d}(r)}{r}
\frac{u_{L'}(R)}{R}
\nonumber \\
& \times \left\{Y_{L'}(\hat{y})\otimes \left\{
Y_{l'}(\hat{x})\otimes \chi_{s'}(1, 2)\right\}_{j'} \right\}_{J'M'}
\nonumber \\
& \times \zeta^{T',0}_{1/2,1/2}(1,2),
\end{align}
where $s'$ and $L'$ are spin and orbital angular momentum of the
entrance channel, respectively, and $l'$ is the orbital angular
momentum of the deuteron. Although in the present study we restrict
ourselves to the $S$-wave component of the deuteron and hence the
quantum numbers $s'=1$ and $l'=0$ are fixed, we aim to derive the
analytical expressions of the matrix elements for a general case of
arbitrary $s'$ and $l'$. In addition,  $u_{l'}^{d}(r)$ is the radial
wave functions of the deuteron and $u_{L'}(R)$ is the scattering
wave function of the $\alpha-d$ pair. The latter asymptotically
behaves as
\begin{equation}
u_{L'}(R) \arrow{R}{\infty}  F_{L'}(k_\omega R) \cos\delta_{L'}(E) +
G_{L'}(k_\omega R) \sin\delta_{L'}(E), \label{eq220}
\end{equation}
where $F_{L'}$ and $G_{L'}$ are Coulomb functions, and
$\delta_{L'}(E)$ is the phase shift in the $L'$-wave at energy $E$.
The parity of the state is defined from the intrinsic parities of
the $\alpha$ particle and deuteron, which are positive and the
orbital momentum $L'$.

The spin and isospin wave functions of the two nucleons as a bound
state of deuteron read, respectively,
\begin{align}
\chi_{s' m'}(1, 2) =\{\chi_{1/2}(1)\otimes \chi_{1/2}( 2)  \}_{s'm'}
\end{align}
and
\begin{align}
\zeta^{T',0}_{1/2,1/2}(1,2)=\{ \zeta_{1/2}(1) \otimes \zeta_{1/2}(2)
\}_{T',0} .
\end{align}
 The antisymmetry condition requires $S'+T'+l'$ to be odd. Since for the deuteron $l'=0$ and $S'=1$, the initial three-body system
 is in the isosinglet state $T'=0$.
The final three-body wave function of the $^6$Li$(1^+,0)$ ground
state in the hyperspherical basis reads as
\begin{align}
 \Psi_{f}^{J M,T0}(\vec{x},\vec{y})=&\frac{1}{\rho^{5/2}}\sum_{\gamma, k}\chi_{\gamma
 k}(\rho)\left\{ {\cal Y}_{l_x l_y}^L (\hat{x},\hat{y})\otimes
 \chi^{S}(\vec{\xi})\right\}_{J M}
 \nonumber \\
& \times
 \Phi_{k}^{l_x l_y}(\alpha) \, \zeta^{T,0}_{1/2,1/2}(1,2),
 \end{align}
where $\rho$ (hyperradius) and $\alpha$ (hyperangle) are defined as
 \beq
 \rho^2 = x^2+y^2, \quad \alpha = \arctan
(y/x). \label{eq203} \eeq Hyperangle $\alpha$ varies between 0 and
$\pi/2$. The hyperspherical harmonics are defined as
\cite{desc03,tur06}
\begin{align}
\Phi_{k}^{l_x l_y}(\alpha)=N_{k}^{l_x l_y}(\cos \alpha)^{l_x}(\sin
\alpha)^{l_y}P_{n}^{l_y+{1}/{2}, l_x+{1}/{2}}(\cos 2\alpha),
\end{align}
where  $P_{n}^{l_y+{1}/{2}, l_x+{1}/{2}}(\cos 2\alpha)$ are the
Jacobi polynomials  and  $N_{k}^{l_x l_y}$  is the normalisation
factor (see Ref.\cite{desc03} for details).

The astrophysical $S$-factor of the process is expressed in terms of
the cross section as \cite{Fowler}
\begin{eqnarray}
S(E)=E \, \sigma_E(\lambda) \exp(2 \pi \eta),
\end{eqnarray}
where  $\eta$ is the Coulomb parameter.

\subsection{Isospin transition matrix elements}

We rewrite the charge operators of the proton and neutron in
Eq.(\ref{eq11}) with the help of the isospin operators as
 \beq
 \hat{Z}_1=\frac{1}{2}+\hat{m}_{t1}, \hspace*{1 cm}
 \hat{Z}_2=\frac{1}{2}+\hat{m}_{t2}.
 \eeq
Then the matrix element of the isospin operator
\begin{align}
\hat{T}_y=& \left[ \left( \frac{1}{2}+\hat{m}_{t1} \right) +\left(
\frac{1}{2}+\hat{m}_{t2} \right) \right] \left(-\frac{A_3}{A}
\right)^{\lambda} \nonumber \\ & + Z_3 \left( \frac{A_{12}}{A}
\right)^{\lambda}
 \end{align}
of the first term in the Eq.(\ref{eq11}) between the initial and
final three-body isospin wave functions reads as
\begin{align}
 \langle \zeta^{T,0}_{1/2,1/2}|\hat{T}_y|\zeta^{T',0}_{1/2,1/2}
 \rangle =
\left[ \left( -\frac{A_3}{A}\right) ^{\lambda} + Z_3 \left(
\frac{A_{12}}{A} \right)^{\lambda} \right] \delta_{T,T'}.
\end{align}
The matrix element of the second isospin operator
\begin{align}
 \hat{T_x}=
 \left( \frac{1}{2}+\hat{m}_{t1} \right) \left(-\frac{A_2}{A_{12}} \right)^{\lambda}
+\left( \frac{1}{2}+\hat{m}_{t2} \right) \left( \frac{A_1}{A_{12}}
\right)^{\lambda}
\end{align}
can be evaluated using the angular momentum algebra
\begin{align}
\nonumber
\langle \zeta^{T,0}_{1/2,1/2}|\hat{T_x}|\zeta^{T',0}_{1/2,1/2}
\rangle =& \frac{1}{2} \left[ \left( -\frac{A_2}{A_{12}}
\right)^{\lambda} + \left( \frac{A_1}{A_{12}} \right)^{\lambda}
\right] \delta_{T,T'} \nonumber \\ & + \frac{1}{2} \left[
\left(-\frac{A_2}{A_{12}} \right)^{\lambda}
-\left(\frac{A_1}{A_{12}} \right)^{\lambda} \right] \nonumber \\
&\times \left(\delta_{T,0} \delta_{T',1}+\delta_{T,1} \delta_{T',0}
\right).
\end{align}
The isospin operator in the last term of Eq.(\ref{eq11}) is
evaluated in the same way as the second term.

From last equation one can note that the E1 transition is allowed
from the
 isospin-singlet states to the
isospin-triplet components of the final $^6$Li$(1^+)$ three-body
bound state. The spin-angular parts of the matrix elements for the
E1- and E2-transition operators in the three-body model are given in
Appendix A.

\section{Numerical results}
\subsection{Details of the calculations}

The radial wave function $u_{l'}^{d}(r)$ of the deuteron  is the
solution of the bound-state Schr{\"o}dinger equation with the
central Minnesota potential $V_{NN}$  \cite{thom77,reich70} with
$\hbar^2/2 m_N=20.7343$ MeV fm$^2$. The Schr{\"o}dinger equation is
solved using a highly accurate Lagrange-Laguerre mesh method
\cite{baye15}. It yields $E_d$=-2.202 MeV for the deuteron
ground-state
 energy with the number of mesh points $N=40$ and a scaling
parameter $h_d=0.40$.

The scattering wave function $u_{L}(E,R)$ of the $\alpha-d$ relative
motion is calculated as a solution of the Schr{\"o}dinger equation
using the Numerov method with an appropriate potential subject to
the boundary condition Eq.(\ref{eq220}). In present study  we use
the well-known deep potential of Dubovichenko \cite{dub94} with a
small modification in the $S$-wave \cite{tur15}:
$V_d^{(S)}(R)=-92.44 \exp(- 0.25 R^2) $ MeV. The potential
parameters in the $^3P_0$, $^3P_1$, $^3P_2$  and $^3D_1$, $^3D_2$,
$^3D_3$ partial waves are the same as in Ref. \cite{dub94}. The
potential contains additional states in the $S$- and $P$-waves
forbidden by the Pauli principle. The above modification allows to
better describe the phase shifts in the $S$-wave, and most
importantly, reproduce the empirical value $C_{\alpha d}=2.31$
fm$^{-1/2}$  of the asymptotic normalization coefficient (ANC) of
the $^6$Li(1+) ground state derived from $\alpha-d$ elastic
scattering data \cite{blok93}.

In order to check the sensitivity of the E1- and E2-transition
matrix elements on the short-range part of the $\alpha-d$ wave
function, we also test the $\alpha-d$ potential $V^S_d$ obtained
from the initial $V_d$ potential in the $S$- and $P$-waves by a
supersymmetric (SUSY) transformation \cite{baye87}. The resulting
potential gives the same phase shifts and the same ground-state
energy as the initial potential. However, the forbidden state is
removed and the role of the Pauli principle is simulated by a
short-range core.

The final $^6$Li(1+) ground-state wave function was calculated using
the hyperspherical Lagrange-mesh method \cite{desc03,tur06,tur07}
with the same Minnesota NN-potential. For the $\alpha-N$ nuclear
interaction the potential of Voronchev et al. \cite{vor168} was
employed, which contains a deep Pauli forbidden state in the
$S$-wave. The potential was slightly renormalized by a scaling
factor 1.008 to reproduce the experimental binding energy $E_b$=3.70
MeV. The Coulomb $\alpha-p$ interaction is parameterized as
$V_C(r)=2e^2 erf(r/R_C)$ with a radius $R_C$=1.2 fm. The Pauli
forbidden states in the three-body configuration space are
eliminated with the help of the orthogonalising pseudopotential
(OPP) method \cite{kuk78,tur98}.

The hypermomentum expansion includes terms up to $K_{\rm max}$ = 20
which ensures a good convergence of the energy. The matter r.m.s.
radius of the ground state (with 1.4 fm for the radius of the
$\alpha$-particle ) was found as
 $\sqrt { \langle r \rangle ^2}$ = 2.25 fm, a value slightly lower than the experimental data (2.32 $\pm$ 0.03 fm \cite{exp162}).
 The ground state is essentially S = 1 (96 percent). As noted above, the three-body wave function includes  also a small isotriplet
 component $l_x=l_y=S=T=1$ with the norm square 1.13 $\times 10^{-5}$ which can give a contribution to the E1-transition matrix elements.

\subsection{Estimation of the astrophysical S-factor}

\begin{figure}[htb]
\includegraphics[width=98mm]{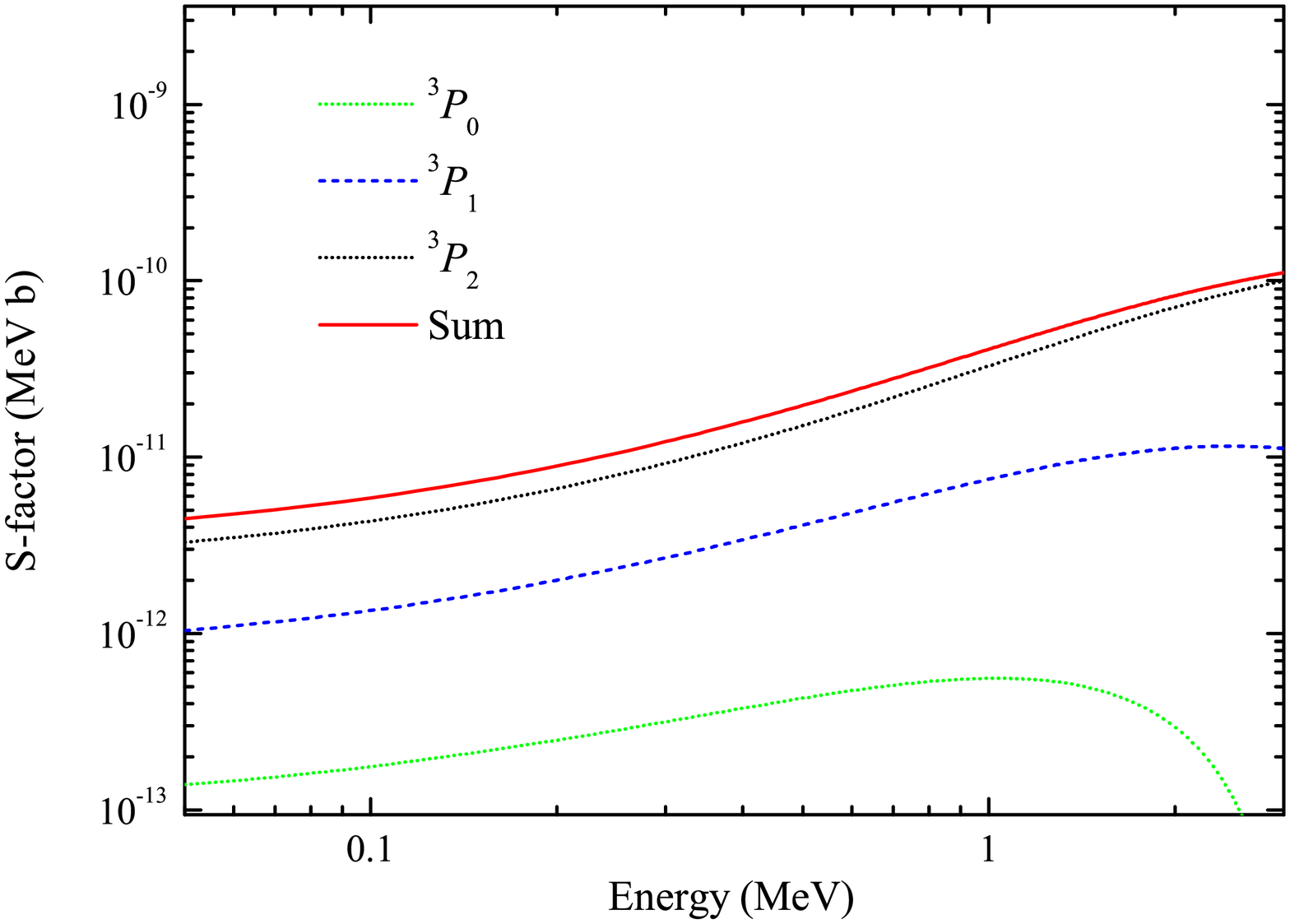}
\caption{Contribution of the E1-transition operator from the initial
isosinglet state to the isotriplet component of the final state for
the astrophysical S-factor of the capture process
$\alpha+d\rightarrow ^6$Li$+\gamma$.   } \label{FIG1}
\end{figure}

\begin{figure}[htb]
\includegraphics[width=98mm]{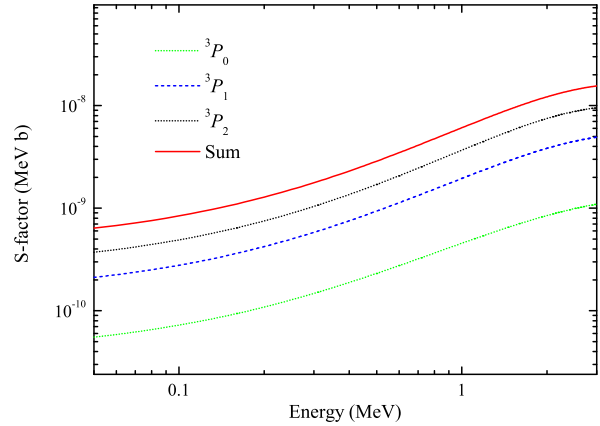}
\caption{Contribution of the E1-transition operator from the initial
isosinglet state to the isotriplet and isosinglet components of the
final state for the astrophysical S-factor of the capture process
$\alpha+d\rightarrow ^6$Li$+\gamma$. } \label{FIG2}
\end{figure}

First we estimate the allowed E1-transition contribution to
 the capture process $^4$He$(d,\gamma)^6$Li in the three-body model when the isospin changes.  Here contributions come from the initial $^3P_0$, $^3P_1$,  $^3P_2$
partial waves and the $l_x=l_y=S=T=1$ components of the final state.
In Fig. \ref{FIG1} we show the corresponding estimation for the
astrophysical S-factor. As can be seen from the picture the
contribution is rather small which means that the small isotriplet
component of the $^6$Li(1+) ground state does not make a significant
contribution to the capture process.
Fig. \ref{FIG2} shows the estimated contribution of the
E1-transition operator to the astrophysical S-factor including the
correction to the mass numbers $A_n$=1.00866491597 a.u., $A_n$=
1.00727646677 a.u. and $A_3$=4.001506179127 a.u. This yields
additional contribution to the S-factor, larger than
isospin-transition terms in Fig. \ref{FIG1} approximately by two
orders of magnitude.

\begin{figure}[htb]
\includegraphics[width=98mm]{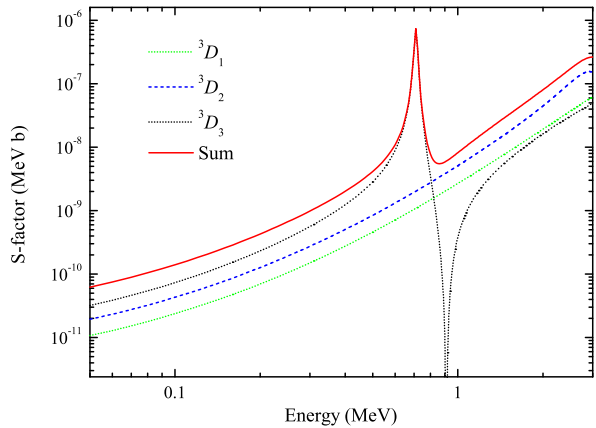}
\caption{Contribution of the E2-transition operator to the
astrophysical S-factor of the capture process $\alpha+d\rightarrow
^6$Li$+\gamma$.} \label{FIG3}
\end{figure}

In Fig. \ref{FIG3} the contribution of the E2-transition operator to
the astrophysical S-factor is demonstrated  for different initial
partial waves $^3D_1$, $^3D_2$ and $^3D_3$.  As can be seen from the
figure the estimations are essentially less than the corresponding
numbers for the two-body model \cite{tur15}.  The magnitude of
underestimation is larger at low astrophysical energies.

 Additionally, unlike the two-body model, in the
three-body model there is a contribution of the initial
$^3S_1$-state to the E2-transition matrix elements. However, our
numerical study shows this contribution to be very small. For the
energy range from 0.1 MeV to 1.0 MeV the $S$-wave contribution to
the astrophysical S-factor increases from 1.$\times10^{-12}$ MeV b
to 2.02$\times10^{-12}$ MeV b. This is why we do not show the S-wave
contribution in Fig. \ref{FIG3}.

\begin{figure}[htb]
\includegraphics[width=98mm]{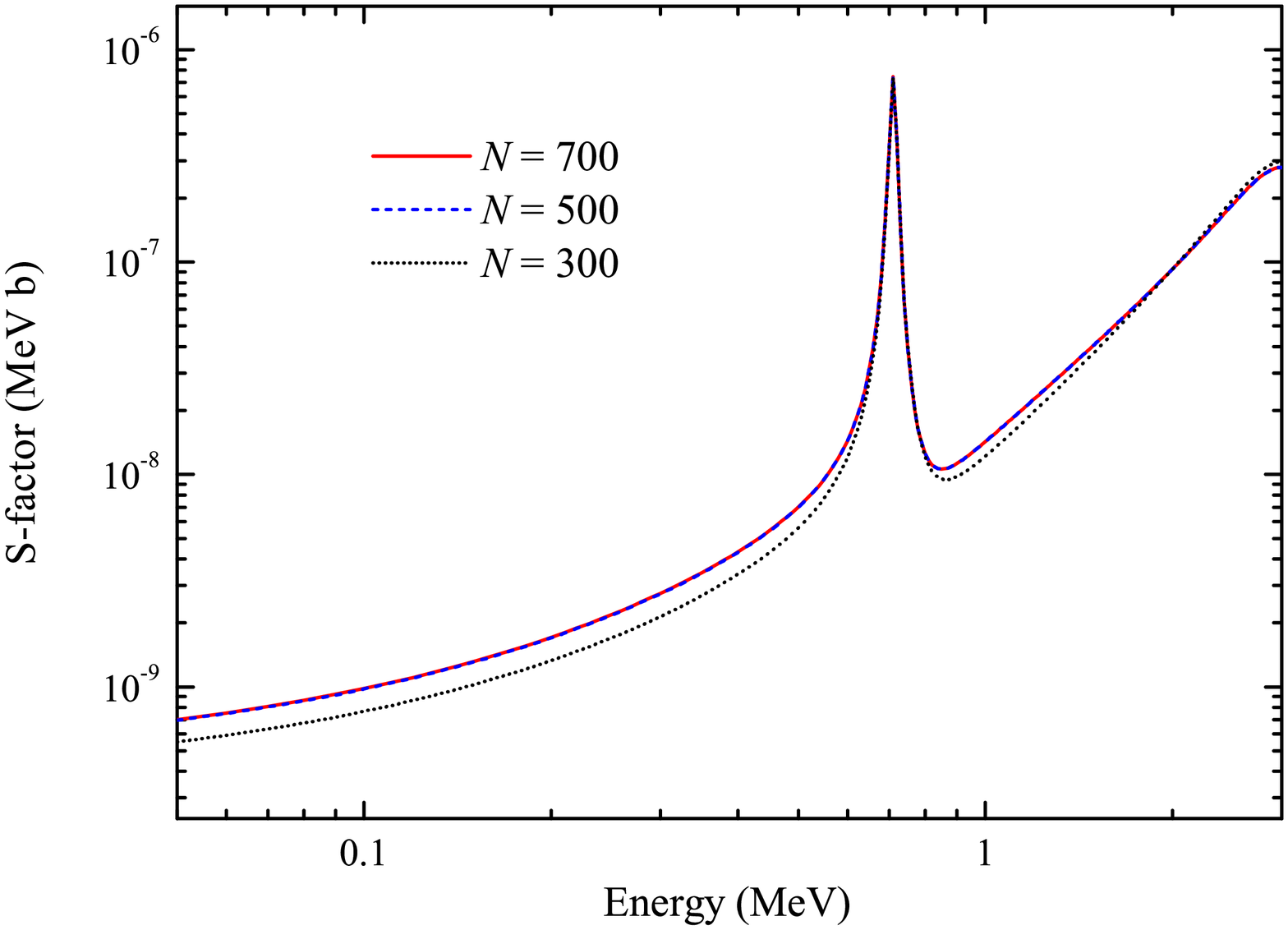}
\caption{Convergence of the astrophysical S-factor for the capture
process $\alpha+d\rightarrow ^6$Li$+\gamma$  with respect to the
number of integration points with the fixed step $h=0.05$ fm.}
\label{FIG4}
\end{figure}

We also have tested the SUSY transformed $V^S_d$ $alpha-d$
potentials. It turns out that this transformation increases the
$S$-wave contribution to the S-factor by about 12-13 percent in the
energy range from 0.1 MeV to 1.0 MeV. But the total $S$-wave
contribution is still negligible. The SUSY transformation of the
$P$-wave potentials yields very small increase of the S-factor by
0.52-0.60 percent in the aforementioned energy range. The situation
is different from the beta- and M1-transition processes
\cite{tur06,tur06a,tur07}, where the main contribution comes from
the $S$-wave $\alpha-d$ scattering state, hence a sensitivity of the
transition probability to the short-range behaviour of the wave
function was essential.

\begin{figure}[htb]
\includegraphics[width=98mm]{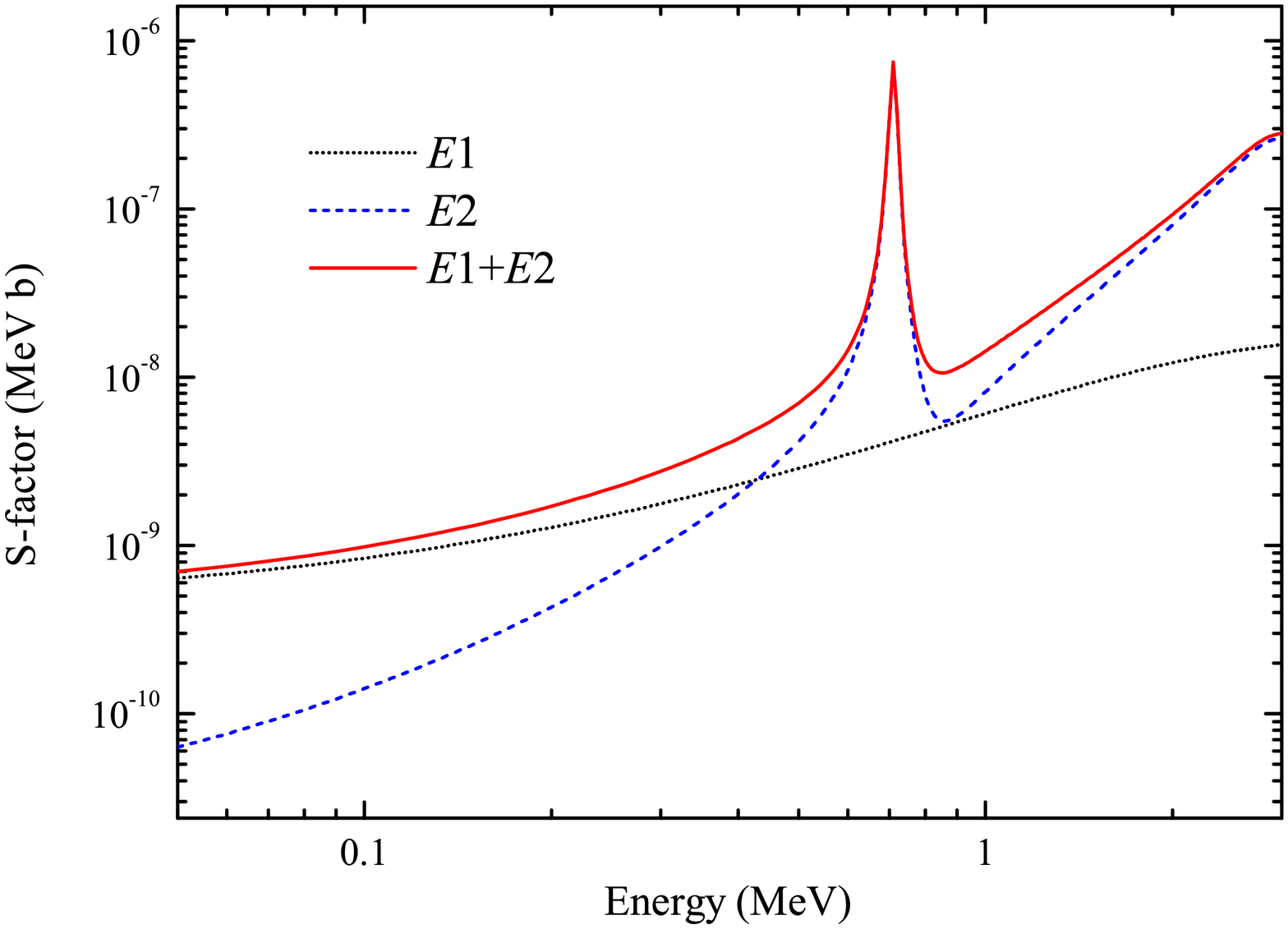}
\caption{Comparison of the contributions of the E1- and
E2-transition operators to the astrophysical S-factor of the capture
process $\alpha+d\rightarrow ^6$Li$+\gamma$.} \label{FIG5}
\end{figure}

Fig. \ref{FIG4} demonstrates the convergence of the evaluated
S-factor in the three-body model  for different choices of the
number of integration
 points $N=300, 500, 700$ with fixed
step $h=0.05$ fm. As one can see, the convergent results are
obtained with N=500 mesh points.
In Fig. \ref{FIG5}  we compare the E1- and E2-transition components.
At low energies the E1 transition dominates and at higher energies
the E2 component is stronger.

\begin{figure}[htb]
\includegraphics[width=98mm]{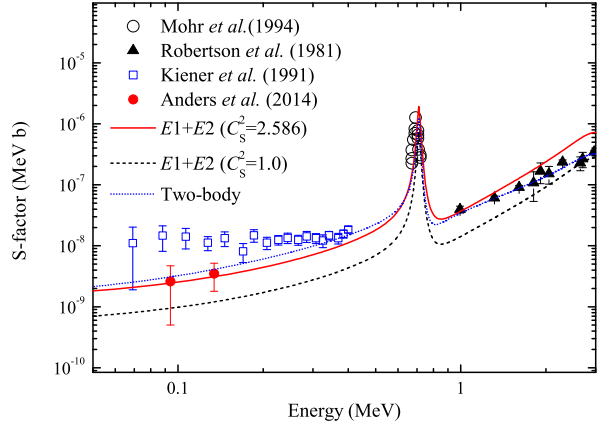}
\caption{Comparison of the theoretical estimations obtained in the
two- and three-body models  for the astrophysical S-factor of the
capture process $\alpha+d\rightarrow ^6$Li$+\gamma$ with available
experimental data.} \label{FIG6}
\end{figure}

Finally, in Fig. \ref{FIG6} we compare the obtained theoretical
results with the estimations of the two-body model \cite{tur15} and
experimental data  from  Refs. \cite{mohr94,robe81,kien91,luna14}.
One can see from the figure, that the results of the two-body and
three-body models differ essentially for the spectroscopic factor
$C_S^2$=1.  At the resonance energy they differ by a factor of 0.565
which is consistent with the square of the overlap integral
$I=0.748$  of the three-body bound state wave function with the
deuteron and the two-body $\alpha-d$ bound state wave functions.

We have estimated the  integral  $P_{\alpha d}=\int
{|\Psi(\vec{R})|^2}d\vec{R} $ with $\Psi(\vec{R})=\langle
\Psi_3(\vec{r},\vec{R}) | \psi_d(\vec{r}) \rangle$ and found its
value to be  0.3867.
 That yields for the spectroscopic factor an estimation $C_S^2=1/P_{\alpha d}$=2.586.
As was shown in Fig. \ref{FIG6} with this value of the spectroscopic
factor the three-body model perfectly describes the new experimental
 data of the LUNA collaboration, better than the two body models. Any value of the spectroscopic
  factor from the interval between 1.50 and 4.25 is able to describe these
 data within the error bar.

\section{Conclusions}

The astrophysical capture process  $\alpha+d\rightarrow
^6$Li$+\gamma$ has been studied in the three-body model.  The
contribution of the E1-transition operator has been estimated from
the initial isosinglet states to the isotriplet components of the
final $^6$Li(1+) bound state. It is shown that this contribution is
small. The most important contribution of the E1 transition comes
due to the mass difference of the proton and neutron with the
violation of the isospin selection rule. The situation is close to
the two-body model where the E1 transition, forbidden by the isospin
selection rule, is only possible due to the mass difference of the
alpha particle and twice the deuteron mass. The three-body model
perfectly matches the new experimental data of the LUNA
collaboration with the spectroscopic factor 2.586 derived from the
overlap integral of the $^6$Li and deuteron bound-state wave
functions.

\acknowledgments

The support of the Australian Research Council, the Australian
National Computer Infrastructure and the Pawsey Supercomputer Centre
are gratefully acknowledged. Authors are thankful to Daniel Baye for
very useful comments. E.M.T. thanks the members of the theoretical
physics group at Curtin University for the kind hospitality during
his visit. A.S.K. acknowledges a partial support from the U.S.
National Science Foundation under Award No. PHY-1415656.

%
\appendix
\section{Spin-angular matrix elements of the $E \lambda$-transition operator in the three-body model}
The spin-angular matrix elements of the $E \lambda$-transition are
given as
\begin{eqnarray}
\langle\psi_{f}^{JM}|M_{\lambda
\mu}^{E}(\vec{x},\vec{y})|\psi_{i}^{J'M'}\rangle&=&\langle
\frac{1}{\rho^{5/2}}\sum_{\gamma, k}\chi_{\gamma k}(\rho)\left\{
Y_{l_x l_y}^L (\hat{x},\hat{y})\otimes
\chi^{S}(\vec{\xi})\right\}_{J M} \Phi_{k}^{l_x l_y}(\alpha)
|M_{\lambda \mu}^{E}(\vec{x},\vec{y})| \nonumber \\ && \times
\frac{u_{l'}^{pn}(r)}{r} \cdot \frac{u_{L'}(R)}{R}
\cdot\left\{Y_{L'}(\hat{y})\otimes \left\{ Y_{l'}(\hat{x})\otimes
\chi_{s'}(1, 2)\right\}_{j'} \right\}_{J'M'}\rangle
\nonumber \\
\end{eqnarray}
where
\begin{eqnarray}
M_{\lambda \mu}^{E}(\vec{x},\vec{y})&=&A_{x}M_{\lambda
\mu}^{E}(\vec{x})+A_{y}M_{\lambda
\mu}^{E}(\vec{y})+\sum_{k>0}^{\lambda-1}A_{x y}^{(k)} \left\{
M_{\lambda-k}^{E}(\vec{x})\otimes M_{k}^{E}(\vec{y})
\right\}_{\lambda \mu}
\end{eqnarray}
and
\begin{eqnarray}
&&\langle \left\{Y_{l_x l_y}^L (\hat{x},\hat{y})\otimes  \chi^{S}(1,
2)\right\}_{J M}
  |A_{x y}^{(k)} \left\{ M_{\lambda-k}^{E}(\vec{x})\otimes M_{k}^{E}(\vec{y}) \right\}_{\lambda \mu}|\left\{Y_{L'}(\hat{y})
  \otimes \left\{ Y_{l'}(\hat{x})\otimes \chi_{s'}(1, 2)\right\}_{j'} \right\}_{J'M'}\rangle
  \nonumber \\
&& \hspace{2cm} = \frac{A_{x y}^{(k)}}{4\pi}\cdot\left(
\frac{x}{\sqrt{\mu_{12}}}\right)^{\lambda-k}\left(
\frac{y}{\sqrt{\mu_{(12)3}}}\right)^{k}\delta_{ss'}[\sigma][\tau]\left(
[k][\lambda-k][\lambda][l'][j'][L'][L][J']\right)^{1/2}
\nonumber \\
&& \hspace{2cm} \times \sum_{\sigma
\tau}(-1)^{2J+2M+l_x+l_y+L-\tau+L'-l'-2\sigma}C^{l_x 0}_{\lambda-k 0
l' 0}C^{l_y 0}_{k 0 L' 0}\left\{ \begin{array}{ccc}
l_y & k & L' \\
l_x & \lambda-k & l'\\
L   & \lambda & \tau
\end{array} \right\}
\left\{ \begin{array}{ccc}
S & L & J \\
l' & \tau & L'\\
j' & \lambda & \sigma
\end{array} \right\}
\nonumber \\ && \hspace{2cm} \times \left\{ \begin{array}{ccc}
\sigma & j' & \lambda \\
J' & J & L'
\end{array} \right\}C^{J M}_{J'M' \lambda \mu}
\end{eqnarray}
\end{document}